\documentclass[12pt]{article}
\usepackage{graphicx}
\usepackage{dcolumn}
\usepackage{bm}

\begin{document}

\begin{center}

{\bf {\LARGE Roper Resonance and $S_{11}(1535)$ from Lattice QCD}}

\vspace{1.5cm}

{\bf N. Mathur$^{a}$, Y. Chen$^{b}$, S.J. Dong$^{a}$, T. Draper$^{a}$, I. Horv\'{a}th$^{a}$, 
F.X. Lee$^{c,d}$, K.F. Liu$^{a}$, and J.B. Zhang$^{e}$}

\begin{flushleft}
{\it $^{a}$Dept.\ of Physics and Astronomy, University of Kentucky, Lexington, KY 40506 \\
$^{b}$Institute of High Energy Physics, Chinese Academy of Science, Beijing 100039, China \\
$^{c}$Center for Nuclear Studies, Dept.\ of Physics, George Washington University, 
Washington, DC 20052 \\
$^{d}$Jefferson Lab, 12000 Jefferson Avenue, Newport News, VA 23606 \\
$^{e}$CSSM and Dept.\ of Physics and Mathematical Physics, University of Adelaide, 
Adelaide, SA 5005, Australia \\
}
\end{flushleft}
\end{center}

\begin{abstract}
Using the constrained curve fitting method and overlap fermions with the lowest
pion mass at $180\,{\rm MeV}$, we observe that the masses of the first positive
and negative parity excited states of the nucleon tend to cross over as the 
quark masses are taken to the chiral limit. Both results at the physical pion mass 
agree with the
experimental values of the Roper resonance ($N^{1/2+}(1440)$) and $S_{11}$
($N^{1/2-}(1535)$).  This is seen for the first time in a lattice QCD
calculation. These results are obtained on a quenched
Iwasaki $16^3 \times 28$ lattice with $a = 0.2\,{\rm fm}$.  We also extract the ghost
$\eta' N$ states (a quenched artifact) which are shown to decouple from the
nucleon interpolation field above $m_{\pi} \sim 300\,{\rm MeV}$.  From the
quark mass dependence of these states in the chiral region, we conclude that
spontaneously broken chiral symmetry dictates the dynamics of light quarks in
the nucleon.
\end{abstract} 

\vfill

The Roper resonance has been studied extensively, but its status as the first
excited state of the nucleon with the same quantum numbers is intriguing.
Although awarded four stars in the Particle Data Table, it is not as easily
identifiable as other resonances.  It does not show up as a peak in the cross
sections in various experiments; rather its existence is revealed through phase
shift analysis~\cite{arn95,hoe93}.  Furthermore, the theoretical situation is
in a quandary.  First of all, it has been noted for a long time that it is
rather unusual to have the first positive parity excited state lower than the
negative parity excited state which is the $N^{1/2-}(1535)$ in the $S_{11}\, \pi
N$ channel.  This is contrary to the excitation pattern in the meson sectors
with either light or heavy quarks.  This parity reversal has caused problems
for the otherwise successful quark models based on $SU(6)$ symmetry with
color-spin interaction between the quarks~\cite{cr00} which cannot accommodate
such a pattern.  Realistic potential calculations with linear and Coulomb
potentials~\cite{lw83} and the relativistic quark model~\cite{ci86} all predict
the Roper to be $\sim$ $100$ -- $200\,{\rm MeV}$ above the experimental value
with the negative parity state lying lower.  On the other hand, the pattern of
parity reversal was readily obtained in the chiral soliton model like the
Skyrme model via the small oscillation approximation to $\pi N$
scattering~\cite{mat87}.  Although the first calculation~\cite{lzb84} of the
original skyrmion gives rise to a breathing mode which is $\sim 200\,{\rm MeV}$
lower than the Roper resonance, it was shown later~\cite{km85} that the
introduction of the sixth order term, which is the zero range approximation for
the $\omega$ meson coupling, changes the compression modulus, yielding better
agreement with experiment for both the mass and width in $\pi N$ scattering.

Since the quark potential model is based on the $SU(6)$ symmetry with residual
color-spin interaction between the quarks, whereas the chiral soliton model is
based on spontaneous broken chiral symmetry, their distinctly different
predictions on the ordering of the positive and negative parity excited states
may be a reflection of different dynamics as a direct consequence of the
respective symmetry.  This possibility has prompted the suggestion~\cite{gr96}
that the parity reversal in the excited nucleon and $\Delta$, in contrast to
that in the excited $\Lambda$ spectrum, is an indication that the inter-quark
interaction of the light quarks is mainly of the flavor-spin nature rather than
the color-spin nature (e.g.\ one-gluon exchange type.) This suggestion is
supported in the lattice QCD study of ``Valence QCD''~\cite{ldd99} where one
finds that the hyperfine splitting between the nucleon and $\Delta$ is largely
diminished when the $Z$-graphs in the quark propagator are eliminated.  This is
an indication that the color-magnetic interaction is not the primary source of
the inter-quark spin-spin interaction for light quarks. (The color-magnetic
part, being spatial in origin, is unaffected by the truncation of $Z$-graphs in
Valence QCD, which only affects the time part.)  Yet, it is consistent with the
Goldstone-boson-exchange picture which requires $Z$-graphs and leads to a
flavor-spin interaction.

The failure of the $SU(6)$ quark model to delineate the Roper and its
photo-production has prompted the speculation that the Roper resonance may be a
hybrid state with excited glue~\cite{bc83} or a $qqqq\bar{q}$ five quark
state~\cite{khk00}.  Thus, unraveling the nature of Roper resonance has direct
bearing on our understanding of the quark structure and chiral dynamics of
baryons, which is one of the primary missions at experimental facilities like 
Jefferson Lab.

Lattice QCD is perhaps the most desirable tool to adjudicate the
theoretical controversy surrounding the issue.  In fact, there have been
several calculations to study the positive-parity excitation of the
nucleon~\cite{lee01,sbo02,ric02,mbb02,ssh02,ehr03}.  However, they have not
been able to probe the relevant low quark mass region while preserving chiral
symmetry at finite lattice spacing (except Ref.~\cite{sbo02} which uses the
Domain Wall fermion).  We employ the overlap fermion~\cite{neu98} on a large
lattice which admits calculations with realistically small quark
masses~\cite{dll00} to study the chiral region.  Since the controversy about
the nature of Roper hinges on chiral symmetry, it is essential to have a
fermion action which explicitly exhibits the correct spontaneously broken
chiral symmetry.

Another difficulty of the calculation of the excited states in lattice QCD is
that the conventional two-exponential fits are not reliable.  Facing the
uncertainty of the fitting procedure for the excited state, it has been
suggested to use a non-standard nucleon interpolating field which vanishes in
the non-relativistic limit~\cite{lee01,sbo02, ric02,mbb02,ehr03}, in the hope
that it may have negligible overlap with the nucleon so that the Roper state
can be seen more readily.  However, the lowest state calculated with this
interpolation field ($2.2\,{\rm GeV}$)~\cite{lee01,sbo02,ric02,mbb02,ehr03} is
much higher than the Roper state.  Employing the maximum entropy method allows
one to study the nucleon and its radial excitation with the standard nucleon
interpolation field~\cite{ssh02}.  However, with the pion mass at $\sim
600\,{\rm MeV}$, the nucleon radial excitation is still too high ($\sim 2\,{\rm
GeV}$).  In order to have a reliable prediction of the excited state in lattice
QCD calculations by way of addressing the above mentioned difficulties, we
implement constrained curve fitting based on an empirical Bayes method.  It has
been advocated~\cite{lcb01,ddh03b} recently as a powerful tool which utilizes
more data points while better controlling the systematics.

\begin{figure}
  \centerline{%
    \includegraphics[width=14.0cm]{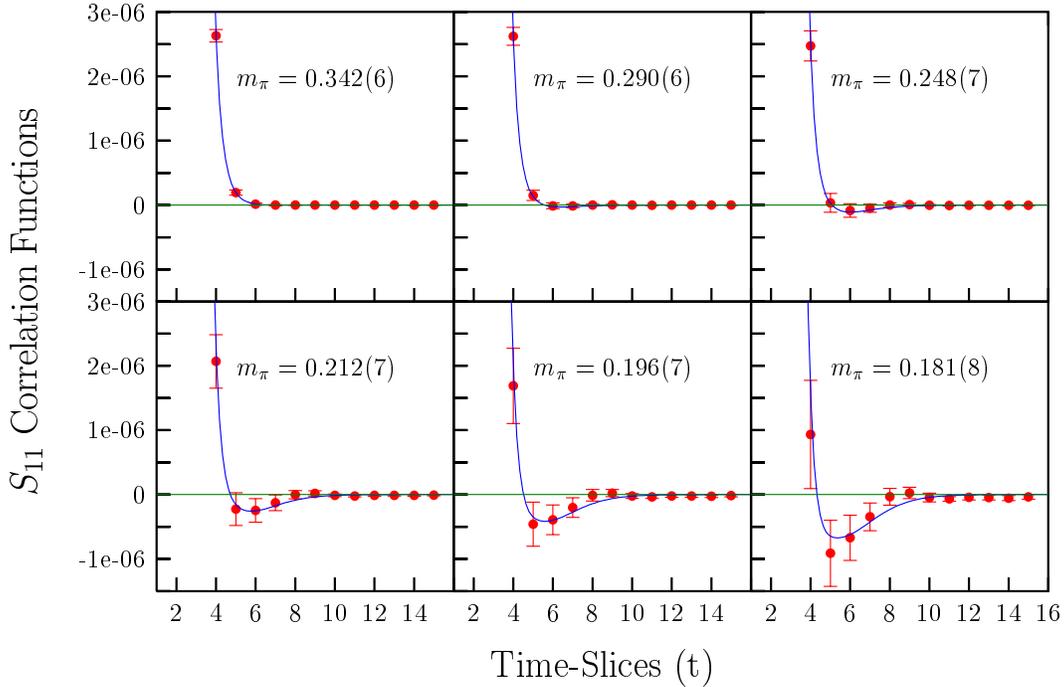}
  }
  \caption{\label{S11_corr} $S_{11}$ correlators for six low quark masses.
    Negative dip of the correlators is an indication of the domination of the
    ghost $\eta^{\prime} N$ state over the physical $S_{11}$.  $m_{\pi}$ are in
    ${\rm GeV}$.  }
\end{figure}

Since we adopt the quenched approximation, we should mention that there are
quenched artifacts.  Due to the absence of quark loops in quenched QCD, the
would-be $\eta'$ propagator involves only double $\eta$ poles in hairpin
diagrams.  This leads to quenched chiral logs in hadron masses which are
clearly observed and extracted from our recent lattice calculation of pion and
nucleon masses~\cite{ddh03a}.  Another quenched artifact is the contribution of
ghost states in the hadron correlators.  As first observed in the $a_0$ meson
channel~\cite{bde02}, the ghost $S$-wave $\eta' \pi$ state lies lower than
$a_0$ for small quark mass.  There is a similar situation with the excited
state of the nucleon where the $P$-wave $\eta' N$ appears in the vicinity of
the Roper.  Since this is not clearly exhibited in the nucleon correlator where
the nucleon is the lowest state in the channel and dominates the long-time
behavior of the correlator, we can look at the parity partner of the nucleon
($N^{\frac{1}{2}-}$ or $S_{11}$) with $I = 1/2$.  There, the lowest $S$-wave
$\eta' N$ state with a mass close to the sum of the pion and nucleon masses can
be lower than $S_{11}$ for sufficiently low quark mass.  Since the $\eta$ --
$\eta$ coupling in the hairpin diagram is negative~\cite{bde02}, one expects
that the $S_{11}$ correlator will turn negative at larger time separations as
in the case of the $a_0$~\cite{bde02}.  In Fig.~\ref{S11_corr}, we show the
$S_{11}$ correlators for 6 low quark cases with pion mass from $m_{\pi} =
181(8)\,{\rm MeV}$ to $m_{\pi} = 342(6)\,{\rm MeV}$.  We see that for pion mass
lower than $248(7)\,{\rm MeV}$, the $S_{11}$ correlator starts to develop a
negative dip at time slices beyond 4, and it is progressively more negative for
smaller quark masses.  This is a clear indication that the ghost $\eta' N$
state in the $S$-wave is dominating the correlator over the physical $S_{11}$
which lies higher in mass.

One can calculate the one-loop hairpin diagram on a $L^3 \times T$ lattice
which gives the following contribution to the zero-momentum nucleon correlator
for the $P$-wave would-be $\eta'$ loop
\begin{eqnarray}  \label{eta-N}
\Delta G_{NN}(t) &=& - m_0^2 \frac{|Z|^2 T}{16 L^3} \sum_{\vec{p}} 
\frac{E_N + E_{\pi} + m_N}{E_N^2\,E_{\pi}^3}\nonumber\\
&&{\hspace*{-0.5in}}\times\, \frac{\vec{p}\,^2}{m_N^2}
 (1 + E_{\pi} t) e^{- (E_N + E_{\pi})t}.
\end{eqnarray}
Here $- m_0^2$ is the coupling between the would-be Goldstone bosons. 
However, this is not quite adequate to be used in the
fitting because there is an interaction between the would-be $\eta'$ and the
nucleon which can lead to an energy which is different from
$E_N + E_{\pi}$.  Instead of modeling it with the re-summation
of the hairpin diagram, as is done for the $a_0$ meson~\cite{bde02}, we shall
introduce these ghost states which are in the vicinity of the Roper and
$S_{11}$ with the form $W (1 + E_{\pi} t) e^{- E_{\eta' N}t}$. This has
the same functional form as in Eq. (\ref{eta-N}). The $(1 + E_{\pi} t)$ pre-factor 
is preserved to reflect the double-pole nature of the hairpin diagram and
the weight $W$ is constrained to be negative to preserve the ghost nature of
the state. The only exception is  $E_{\eta' N}$ which is allowed to be different
from but close to the non-interacting energy $E_N + E_{\pi}$. Being the energy of 
the interacting would-be $\eta'$ and $N$ state, it will be fitted to the data together with
the negative weight $W$.
Since we work in a finite box, the $\eta' N$ states are discrete and they will
be constrained to be near the energy of the two non-interacting particles,
with discrete lattice momenta $p_n = \frac{2\pi n}{L}, n = 0, \pm 1, \cdots$.

%
%

With a constrained-curve fitting method~\cite{ddh03b}, we develop an adaptive
sequential way of extracting the priors from data at a smaller volume, i.e.\
$12^3 \times 28$.  We fit the ground state first over a few large-time slices,
and then use the fitted values as priors to constrain the ground state in a
subsequent fit (which includes earlier time slices) of the excited
state.  We include the next excited states, one by one, constraining all
previously fitted states to their most recently fitted values.  For each new
fit, we add one more earlier time slice and check if it is needed by 
comparing its contribution to the statistical error of the added
time slice. When the time slices are exhausted, we
use the resultant fit parameters as priors for a standard constrained-curve fit
of the correlator on the larger $16^3 \times 28$ lattice.  
More details can be found in Ref.~\cite{ddh03b}. For the $\eta' N$
state, we make an appropriate adjustment to priors for $E_{\eta' N}$ and
$E_{\pi}$ due to the fact that the discrete momentum is inversely proportional
to the spatial lattice size $L$.  We find that the constraint to make $W$
negative is crucial to obtain a fit with acceptable $\chi^2$, especially for
the case of the nucleon correlator where the Roper and the first $\eta' N$
state are nearly degenerate.

\begin{figure}[t]
\vspace*{-0.3in}
  \centerline{%
    \includegraphics[width=0.58\hsize]{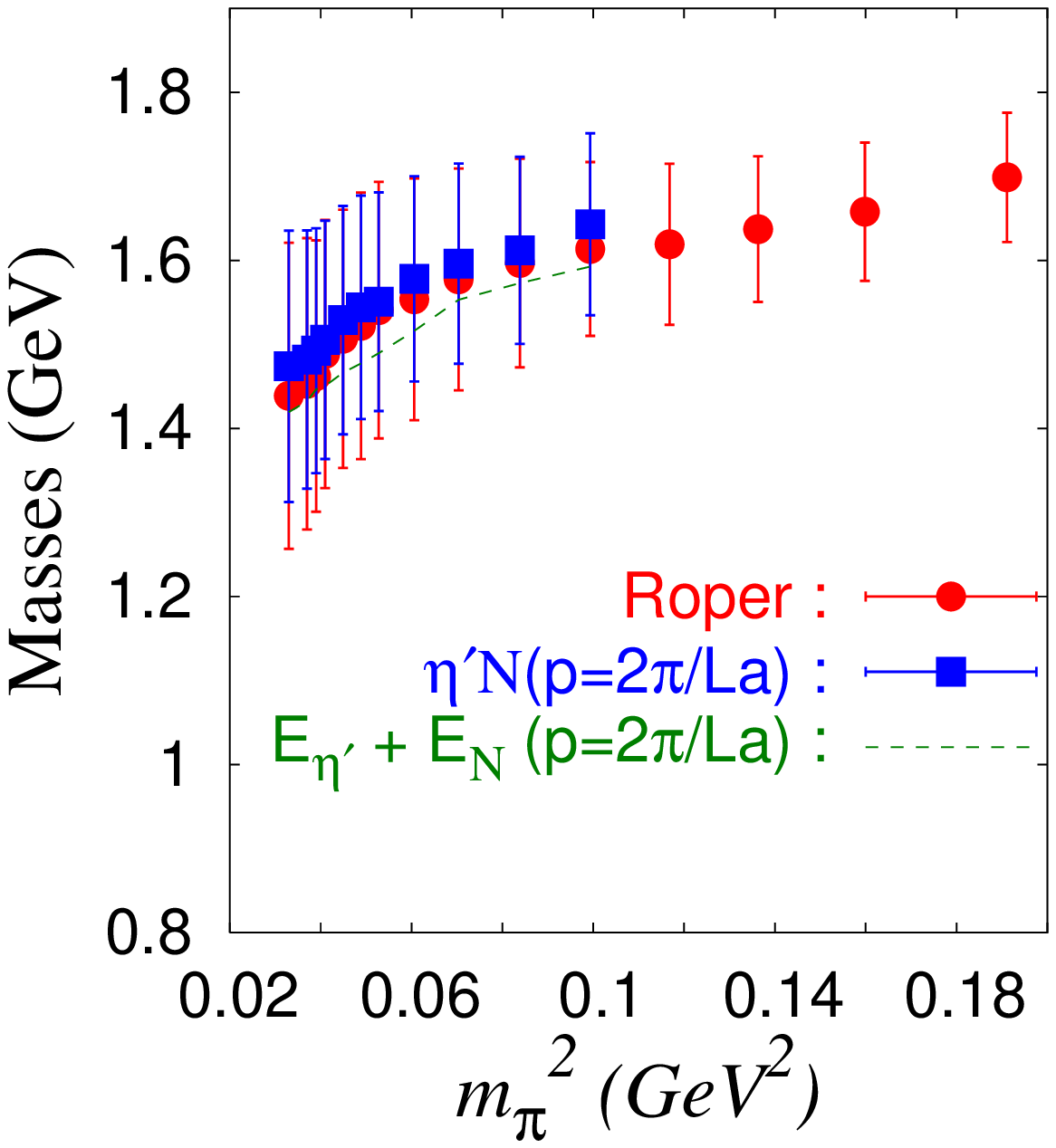}
    \hspace*{-0.16\hsize}
    \includegraphics[width=0.58\hsize]{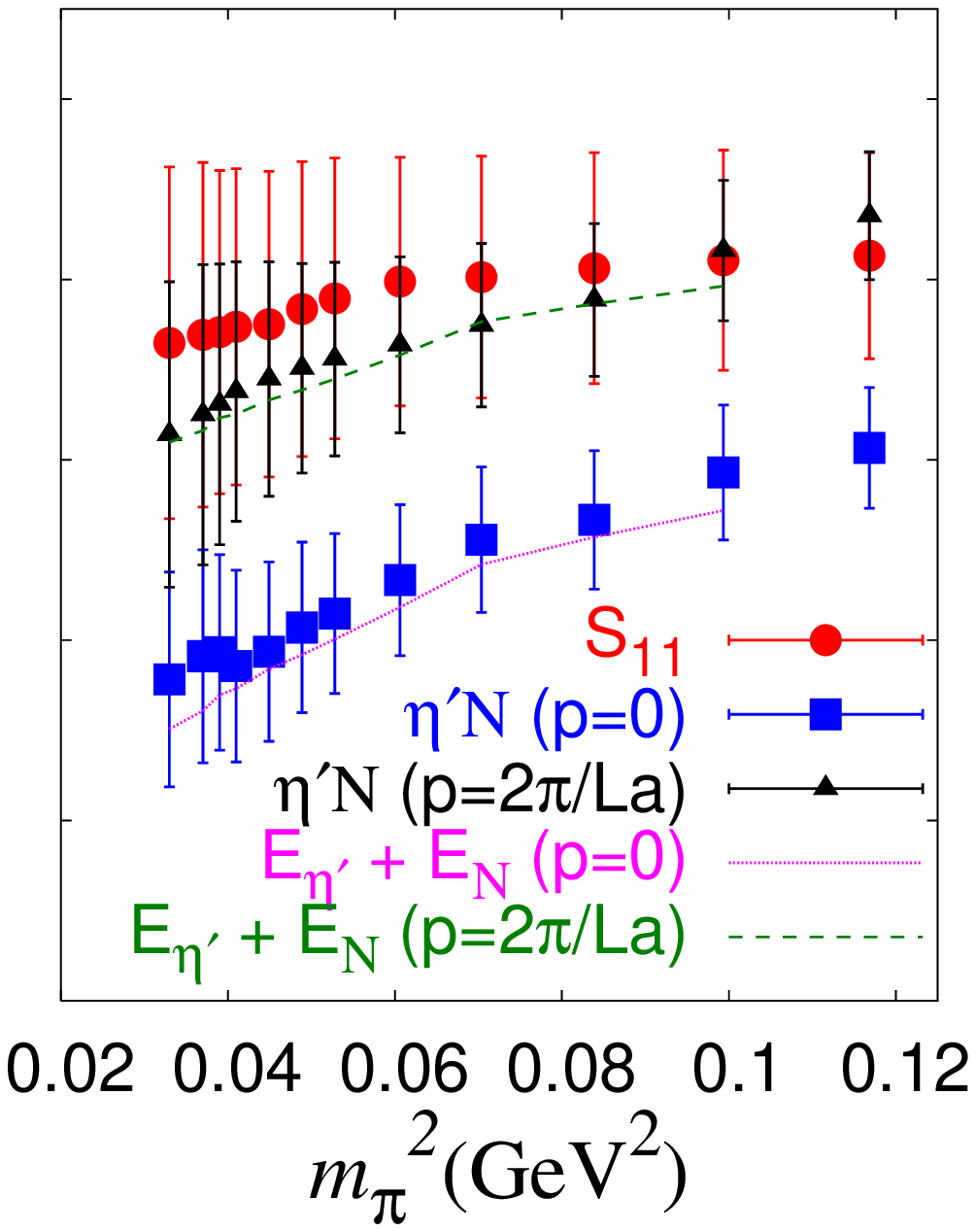}
  }
  \caption{\label{S_11_mass}
    Roper (left) and $S_{11}$ (right) masses with their associated
    $\eta^{\prime} N$ ghost states.  
  }
\end{figure}

We shall report our results on $16^3 \times 28$ and $12^3 \times 28$ lattices
with quenched Iwasaki gauge action and overlap fermions with a lattice spacing
$a = 0.200(3)\,{\rm fm}$ determined from $f_{\pi}(m_{\pi})$.  This makes our
lattice sizes as large as $3.2\,{\rm fm}$ and $2.4\,{\rm fm}$.  At our lowest
pion mass at $181(8)\,{\rm MeV}$, the finite volume error is estimated to be
$\sim 2.7\%$~\cite{ddh03a}.  We first present the Roper and $S_{11}$ masses for
the $16^3 \times 28$ lattice with their associated ghost $\eta' N$ states in
the $P$-wave and $S$-wave respectively.  They are plotted as a function of
$m_{\pi}^2$ in Fig.~\ref{S_11_mass}.  We see in Fig.~\ref{S_11_mass} (left)
that the $\eta' N$ state ($p = 2\pi/L$) and the Roper are nearly degenerate;
they are resolved when the weight for the $\eta' N$ is constrained to be
negative.  Fig.~\ref{S_11_mass} (right) shows that the $p =0$ $S$-wave $\eta'
N$ lies lower than $S_{11}$, and below $m_{\pi} \sim 316\, {\rm MeV}$ the $p =
2\pi/L$ $\eta' N$ state also becomes lower than $S_{11}$.  The dashed line in
Fig.~\ref{S_11_mass} (left) is the mass for the non-interacting would-be
$\eta'$ and $N$  which lies below the fit of the $\eta'
N$ state.  This shows that there is a repulsive interaction of about $50\, {\rm MeV}$ 
between the would-be $\eta'$ and $N$ in the $P$-wave.  The
corresponding non-interacting $\eta'N$ states, represented by the solid line ($p
= 0$) and the dashed line ($p = 2\pi/L$) in Fig.~\ref{S_11_mass} (right), are
closer to the interacting $\eta' N$ states with $\sim 10 - 50\,{\rm MeV}$
repulsion in this case. 

\begin{figure}[t]
\vspace*{-0.1in}
  \centerline{%
    \includegraphics[width=0.57\hsize]{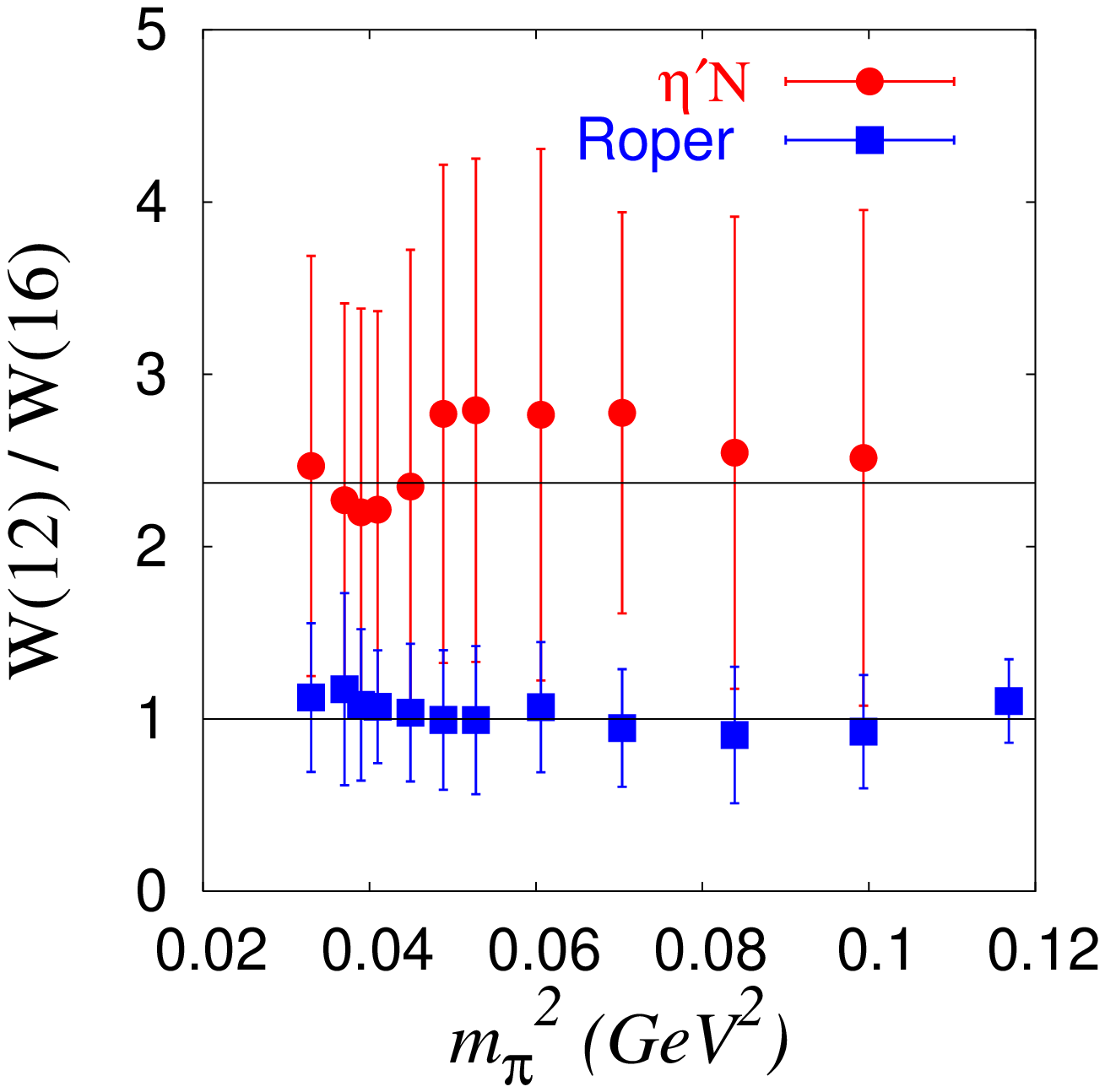}
    \hspace*{-0.13\hsize}
    \includegraphics[width=0.57\hsize]{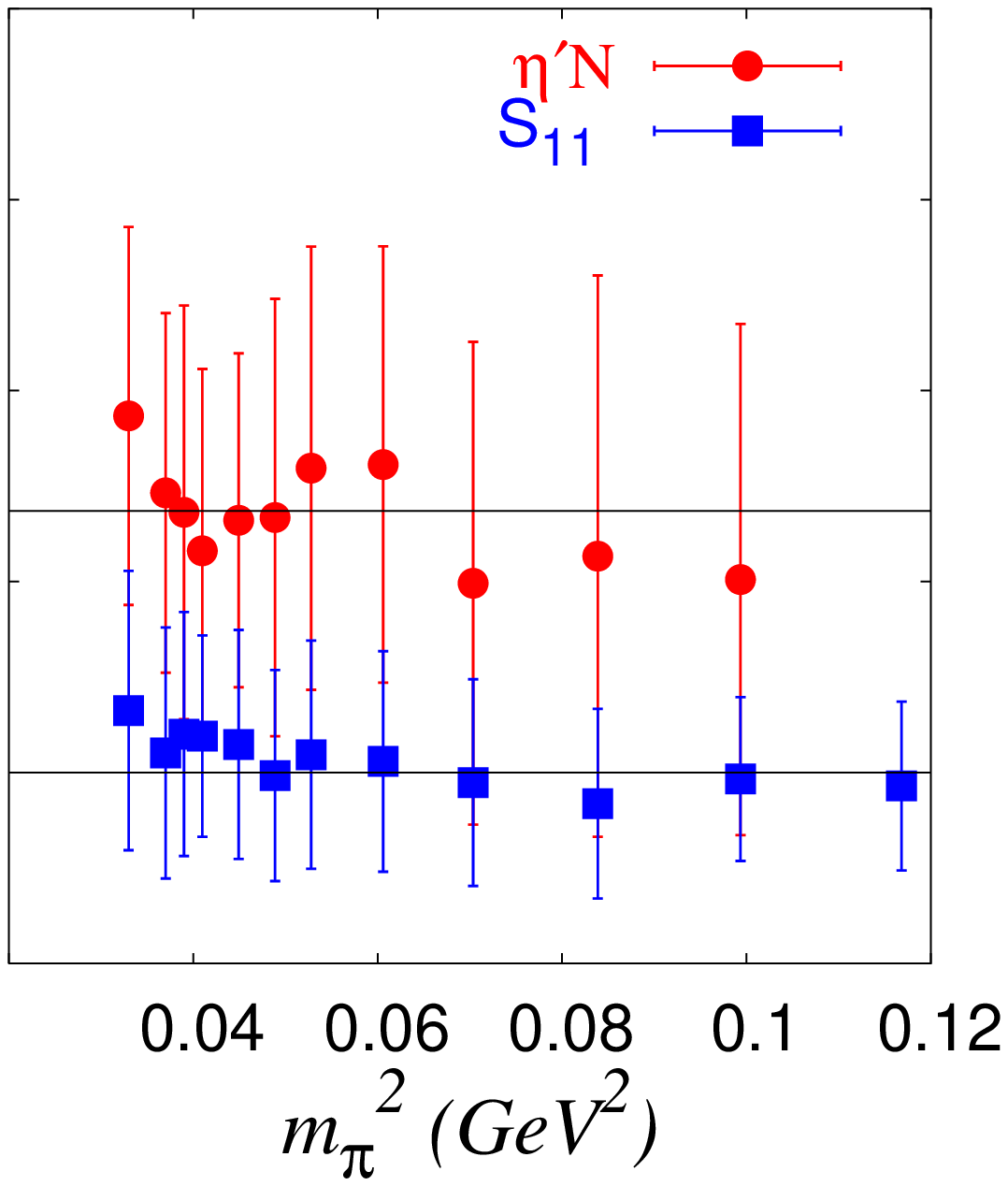}
  }
  \caption{\label{ratio}
    Ratio of spectral weights from $12^{3}$ to $16^{3}$ lattices for Roper and $P$-wave
    $\eta^{\prime} N$ (left) and $S_{11}$ and $S$-wave $\eta^{\prime} N$
    (right).  
  }
\end{figure}

To verify that the $\eta' N$ state can be separated from the physical Roper and
$S_{11}$, we examine the ratio of the spectral weights for the Roper (respectively, 
the $S_{11}$) and
the $\eta' N$ from the $12^3 \times 28$ lattice to that on the $16^3 \times 28$
lattice.  (See Fig.~\ref{ratio}.)  The ratio for the Roper ($S_{11}$) is consistent
with unity which is expected for a one-particle state; whereas, 
the corresponding ratios for the $\eta' N$ states are more consistent with
the ratio of the three volumes $V_3(16)/V_3(12) = 2.37$ which reflects the
volume dependence of the two-particle scattering state.

\begin{figure}[t]
\vspace*{-0.1in}
  \centerline{%
    \hspace*{-0.1\hsize}
    \includegraphics[width=0.65\hsize]{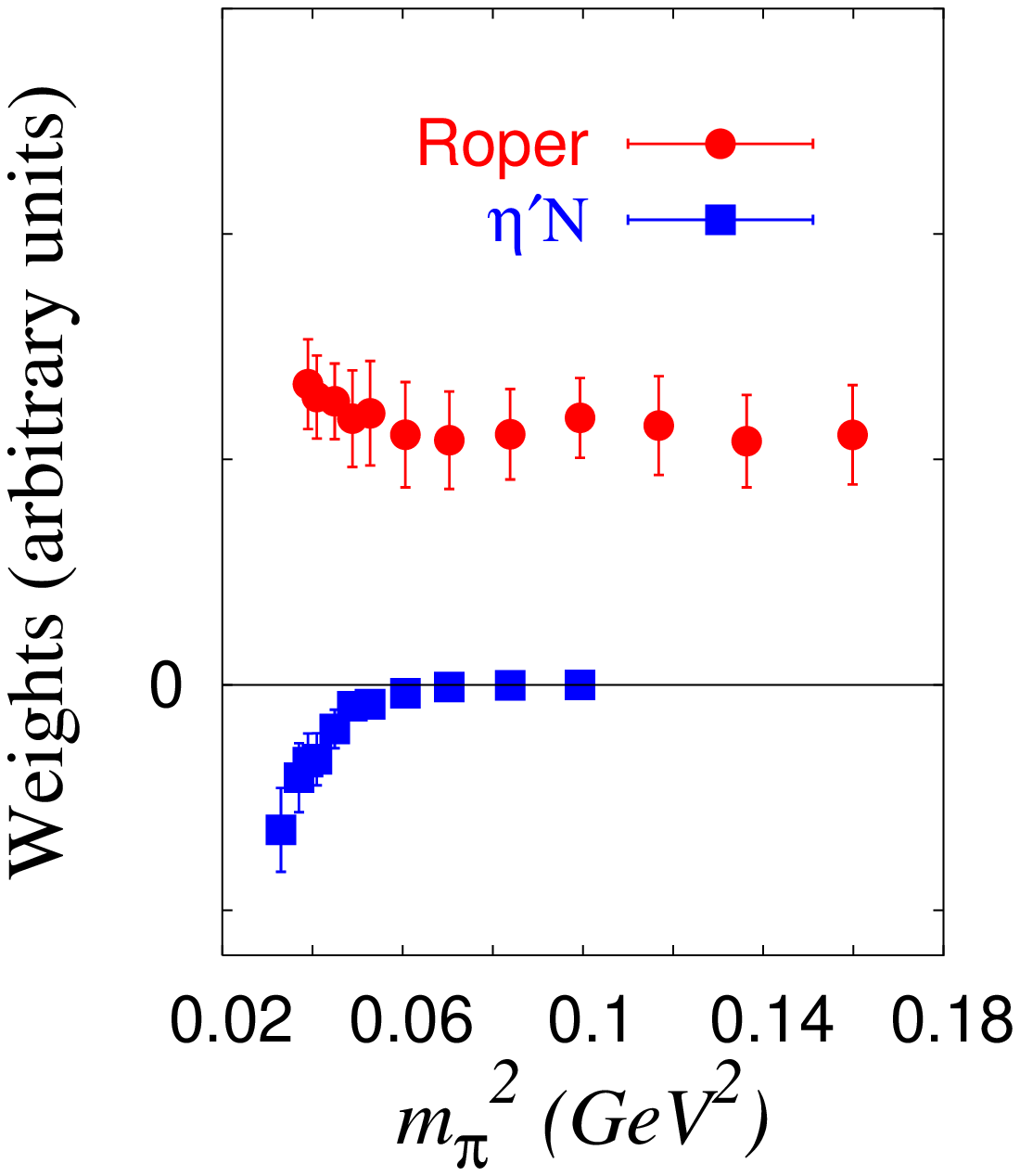}
    \hspace*{-0.2\hsize}
    \includegraphics[width=0.65\hsize]{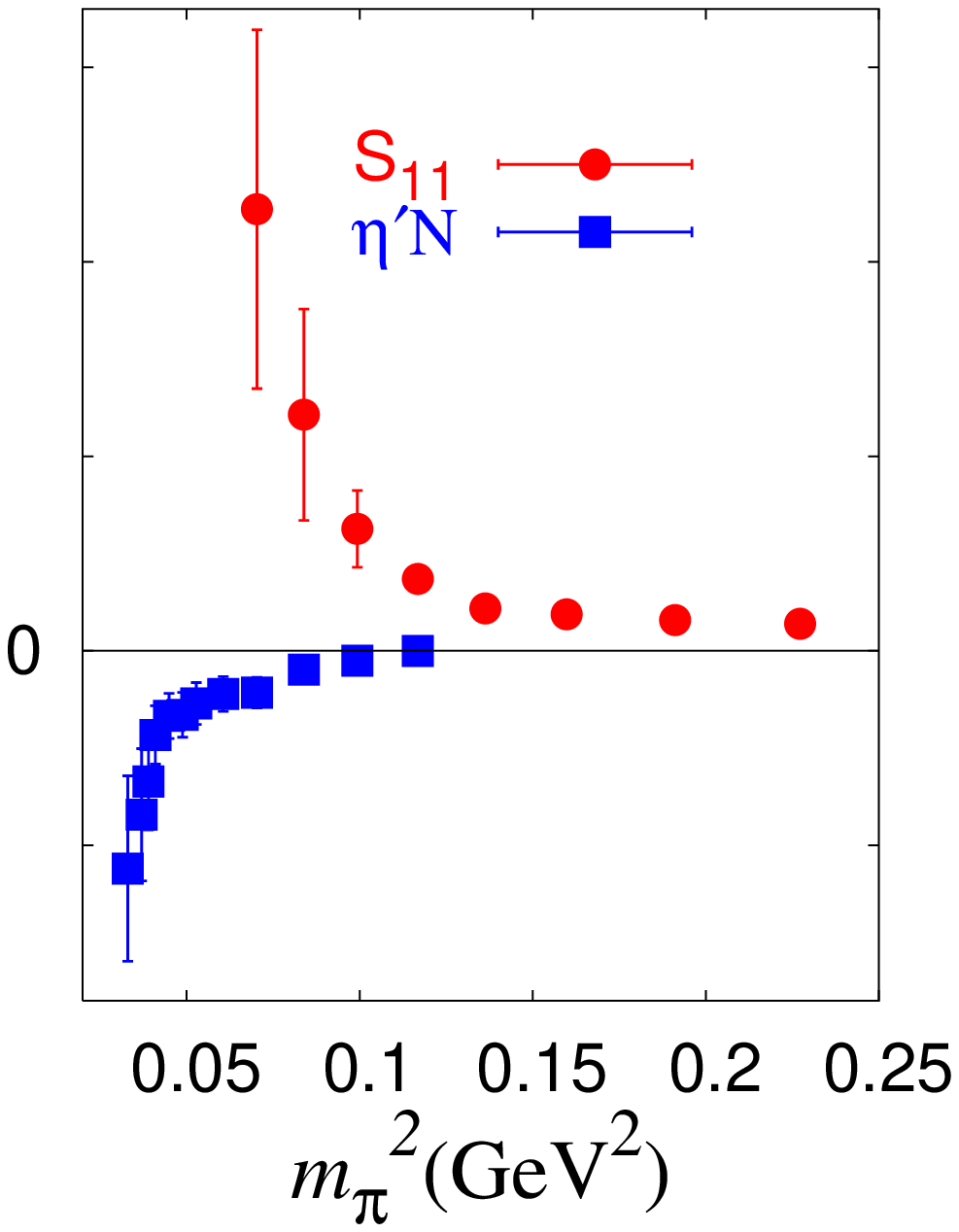}
  }
  \caption{\label{weight}
    Weights of Roper and $\eta^{\prime} N$ (left), and $S_{11}$ and
    $\eta^{\prime} N$ (right).  They are in arbitrary units.  
  }
\end{figure}

In Fig.~\ref{weight}, we display the weight of the $\eta' N$ state and those of
the Roper and $S_{11}$.  The weight for the Roper is essentially constant
(Fig.~\ref{weight} (left)) and that of $S_{11}$ decreases as the pion mass
increases and reaches a constant for $m_{\pi} > 400\, {\rm MeV}$
(Fig.~\ref{weight} (right)).  The negative weight of the $\eta' N$ is 3 orders of
magnitude smaller than the Roper above $m_{\pi} = 266\, {\rm MeV}$ in the nucleon channel 
and $m_{\pi} = 320\, {\rm MeV}$ in the $S_{11}$ channel.  This decoupling means that the
two-particle intermediate state $\eta' N$ is not produced by the three-quark
nucleon interpolating field $\chi_N^{\dagger}$ so that the matrix element
$\langle \eta' N|\chi_N^{\dagger}|0\rangle$ is zero.  This confirms our earlier
study of Valence QCD~\cite{ldd99} which suggests that, for higher quark
mass, the $SU(6)$ symmetry prevails which explains why the valence
quark model with color-spin interaction works well.  However, for light quarks,
it is chiral symmetry which dictates the dynamics via the $Z$-graphs.  This is
to be manifested as the meson loops as well as meson exchanges between quarks
in the chiral quark model of the baryon~\cite{ldd00}.  The present study
suggests that this transition occurs in the region $m_{\pi} \sim 300\,{\rm
MeV}$. We should point out that the fact that the $\eta' N$ ghost states are separated out
from the physical $S_{11}(1535)$ in our fitting relies on the fact that the ghost state
has a different functional form from that of the physical state and its weight is
negative. When the dynamical fermion calculation is available with realistically light quarks,
these $\eta' N$ ghost states will turn into physical $\eta N$ and $\eta' N$ states.
Since $S_{11}(1535)$ is close to the $\eta N$ threshold and both have the
same functional form, i.e. $W\,e^{-mt}$, it would be difficult to separate them with
a fitting procedure based on one interpolation field.

\begin{figure}[t]
\vspace*{-0.2in}
  \centerline{%
    \includegraphics[width=14.0cm]{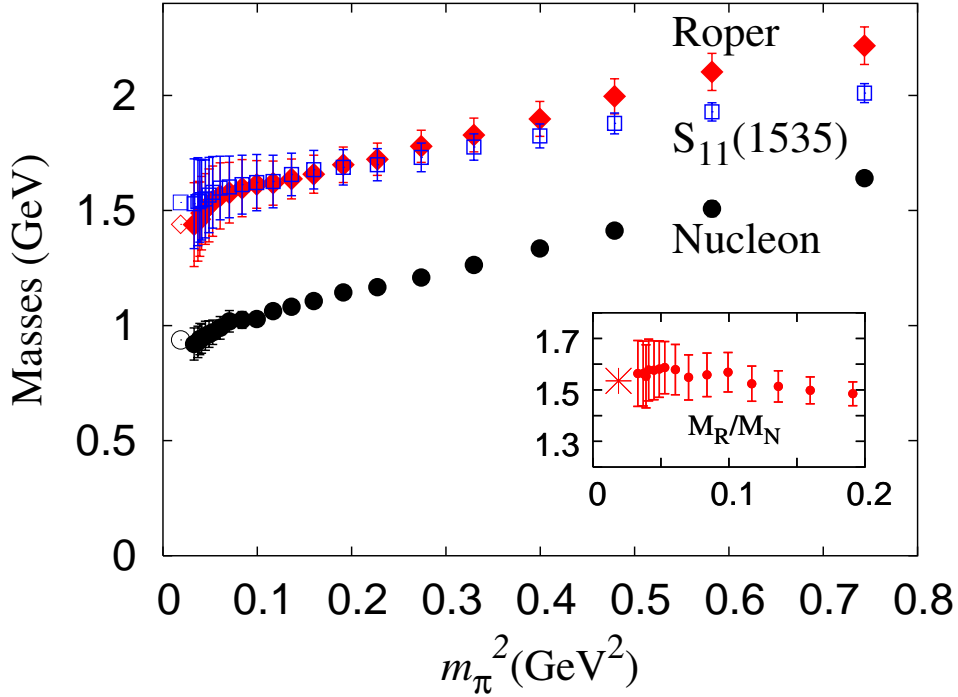}
  }
  \caption{\label{nu_R_S11} 
    Nucleon, Roper, and $S_{11}$ masses in ${\rm GeV}$ as a function of
    $m_{\pi}^2$.  The experimental values are indicated by the corresponding
    open symbols.  }
\vspace*{-0.08in}
\end{figure}

Finally, we plot the nucleon, Roper and $S_{11}$ in Fig.~\ref{nu_R_S11} as a
function of $m_{\pi}^2$.  We see that for heavy quarks ($m_{\pi} \ge 800$ MeV),
the Roper, $S_{11}$, and nucleon splittings are like those of the heavy quarkonium.
When the quark mass becomes lighter, the Roper and $S_{11}$ have a tendency to 
coincide and possibly cross over. More statistics are needed to clarify this point.
However, we have plotted as the insert in Fig.~\ref{nu_R_S11}
the ratio of Roper to nucleon as a function of $m_{\pi}^2 (\rm{GeV}^2)$ which
has a smaller error (by $\sim 40\%$) than the Roper mass itself. It shows that
the Roper is consistent with the experimental value near the physical pion mass. 
The chiral
behavior of the nucleon mass is analyzed in more detail with the inclusion of
the $m_{\pi} \log (m_{\pi})$ term in~\cite{ddh03a}.  Here, we shall simply use
the form $C_0 + C_{1/2} m_{\pi} + C_1 m_{\pi}^2$ to extrapolate the Roper and
$S_{11}$ to the physical pion mass.  The resultant nucleon mass of $928(56)\,{\rm
MeV}$, Roper at $1462(157)\,{\rm MeV}$, and $S_{11}$ at $1554(65)\,{\rm MeV}$
are consistent with the experimental values ($m_N = 939\,{\rm MeV}$, $m_R
\approx 1440\,{\rm MeV}$, and $m_{S_{11}} \approx 1535\,{\rm MeV}$).
This suggests that a successful description of the Roper resonance depends not so 
much on the use of the dynamical quarks, but that most of the essential
physics is captured by using light quarks to ensure the correct chiral behavior.

To conclude, with the help of constrained curve fitting (which allows fitting
of the excited states including the ghost $\eta' N$ states) and the overlap
fermion (which admits calculations with realistically small quark masses), we
are able to study the Roper and $S_{11}$ closer to the chiral limit than before.  The
physical Roper and $S_{11}$ are distinguished from the ghost two-particle
intermediate states ($\eta' N$) by checking their volume dependence of the 
weights as a function of the pion mass. We observe the tendency for the Roper
and $S_{11}$ to cross over (with the caveat of large errors) at the low $m_{\pi}$
region. This is consistent with the thesis that the order reversal between the Roper and
$S_{11}(1535)$ compared to the heavy quark system is caused by the flavor-spin
interaction between the quarks due to Goldstone boson exchanges~\cite{gr96}.
We support the notion that there is a transition from heavy quarks (where the
$SU(6)$ symmetry supplemented with color-spin interaction for the valence
quarks gives a reasonable description) to light quarks (where the dynamics is
dictated by chiral symmetry).  It is suggested that this transition occurs at
$m_{\pi} \sim 300-400\,{\rm MeV}$ for the nucleon.

This work is partially supported by DOE Grants DE-FG05-84ER40154 and
DE-FG02-95ER40907. We wish to thank A. Bernstein, G.E. Brown, L. Glozman, 
D.O. Riska, and S. Sasaki for useful discussions.

\end{document}